\documentclass[11pt]{article}
\usepackage{amsmath,amsfonts,amssymb,amsthm,thmtools}
\usepackage[colorlinks=true, allcolors=blue]{hyperref}
\usepackage[a4paper,left=2.9cm,right=2.9cm,top=3cm,bottom=3cm]{geometry}
\usepackage{xcolor}
\usepackage{mathrsfs}
\usepackage{verbatim}
\usepackage{authblk}
\usepackage[round,authoryear]{natbib}
\usepackage{cleveref}
\usepackage[normalem]{ulem}

\newtheoremstyle{thmstyle}%
{}{}
{\normalfont}{} 
{\bfseries}{.} 
{ }{\thmname{#1}\thmnumber{ #2}\thmnote{\normalfont\ (#3)\bfseries}}

\newtheoremstyle{assstyle}%
{}{}
{\itshape}{} 
{\bfseries}{.} 
{ }{\thmname{#1}\thmnumber{ #2}\thmnote{\normalfont\ (#3)}}

\theoremstyle{thmstyle}
\newtheorem{theorem}{Theorem}
\newtheorem{proposition}{Proposition}
\newtheorem{lemma}{Lemma}
\newtheorem{remark}{Remark}
\newtheorem{corollary}{Corollary}
\newtheorem{example}{Example}

\newtheorem*{claim*}{Claim}

\theoremstyle{assstyle}
\newtheorem{assumption}{Assumption}
\crefname{assumption}{Assumption}{Assumptions}
\Crefname{assumption}{Assumption}{Assumptions}

\newcommand{\sX}{\mathsf{X}}
\newcommand{\ind}{\mathbf{1}}
\newcommand{\cB}{\mathcal{B}}
\newcommand{\cE}{\mathcal{E}}
\newcommand{\cH}{\mathcal{H}}
\newcommand{\cN}{\mathcal{N}}
\newcommand{\cL}{\mathcal{L}}
\newcommand{\cF}{\mathcal{F}}
\newcommand{\bR}{\mathbb{R}}
\newcommand{\bN}{\mathbb{N}}

\newcommand{\rR}{\mathrm{R}}
\newcommand{\rL}{\mathrm{L}}
\newcommand{\rI}{\mathrm{I}}
\newcommand{\ff}{\tilde{f}}
\newcommand{\gap}{\operatorname{Gap}}
\newcommand{\osc}{\operatorname{osc}}
\newcommand{\var}{\operatorname{Var}}
\newcommand{\tr}{\operatorname{tr}}
\newcommand{\argmin}{\operatorname{argmin}}

\providecommand{\keywords}[1]{%
    \textbf{Keywords:} #1
}

\title{Spectral Gap Inheritance for Metropolis--Hastings with Reversible Proposals}

\author{Minhyeok Kim}
\author{Vinayak Rao}
\affil{Department of Statistics, Purdue University}

\begin{document}

\maketitle

\begin{abstract}
We study Metropolis--Hastings kernels whose proposal kernels are reversible with respect to a reference measure, and establish sufficient conditions for such kernels to inherit a nonzero right spectral gap from their proposal kernel. Our conditions require the importance weight of the target relative to the reference measure to be bounded above, but do not require it to be bounded away from zero. Our approach combines a localized Holley--Stroock argument, which yields a local Poincar\'e inequality, with a Lyapunov drift condition. As our main application, we analyze the preconditioned Crank--Nicolson (pCN) algorithm for sampling from measures on a Hilbert space that are absolutely continuous with respect to a Gaussian reference measure. Under local boundedness and {quasiconvexity} assumptions on the potential, we establish a dimension-independent $L^2$-spectral gap for this Metropolis--Hastings algorithm, thereby covering cases in which the local Lipschitz condition used by~\citet{hairer14gappcn} is not satisfied.
\end{abstract}


\section{Introduction}\label{sec:intro}

Markov chain Monte Carlo (MCMC) methods provide a flexible framework for simulation from complex probability distributions. Used widely in Bayesian statistics, these methods involve constructing a Markov chain 
that, under suitable regularity conditions, converges in distribution to an intractable target distribution. 
Geometric ergodicity (GE) and spectral gap estimates are central tools for quantifying the convergence and efficiency of MCMC methods. In particular, such convergence properties, together with suitable moment conditions, yield central limit theorems for Monte Carlo averages and hence provide theoretical guarantees for MCMC estimation~\citep{roberts97ge,chan_94_MC_posterior,flegal_08_MC_error}. 
For reversible Markov chains, these convergence properties can be studied through functional inequalities such as the Poincar\'e inequality, which characterizes the right spectral gap of the Markov kernel.

The Independent Metropolis--Hastings (IMH) algorithm is a popular MCMC algorithm for which these questions are particularly well understood. This is a Metropolis--Hastings (MH) algorithm~\citep{metropolis53,hastings70} where a new state is proposed independent of the current state of the chain. It commonly forms a component of Metropolis-within-Gibbs~\citep{lee_18_scaling_imh} and sequential Monte Carlo samplers~\citep{south_19_smc_imh}. IMH is known to be geometrically ergodic if and only if the target-to-proposal importance weight is essentially bounded above~\citep{mengersen96mh}, with~\citet{wang22exact} characterizing its exact convergence rate.

In this work we consider a broader class of Metropolis--Hastings algorithms in which proposals are generated by a reversible Markov kernel that has a stationary distribution.
When the Markov kernel is just the stationary distribution, then this reduces to the IMH algorithm. Generalizations beyond independent proposals are useful in two complementary settings:
\begin{description}
    \item[Cheap but poorly matched proposals] Independent samples from a poorly matched proposal distribution can lead to a high rejection rate, so that a chain that is nominally geometrically ergodic can still mix poorly in practice. 
    A canonical example is IMH in Bayesian inference with the prior as proposal: the posterior is typically much more concentrated than the prior, so most independent draws from the prior are rejected. 
    Replacing independent sampling with a Markov kernel that preserves the proposal distribution keeps proposals close to the current state, and when appropriately designed, such local moves can substantially improve mixing relative to IMH. The preconditioned Crank--Nicolson (pCN) algorithm~\citep{cotter13mcmc} is an example of this construction: it replaces independent draws from a Gaussian prior with an autoregressive update that leaves the prior invariant.
    \item[Well-matched but expensive proposals] In this setting, a Markov kernel targeting the proposal distribution can make it possible to exploit a proposal that would otherwise be computationally impractical. For example, \citet{pei26gibbsea} considers a target distribution that can be viewed as a Gaussian process (GP) posterior weighted by a term depending on the extrema of the process. One natural approach is to use the GP posterior as a proposal distribution, with the additional weighting term determining a simple acceptance probability. However, when the likelihood is non-Gaussian, direct sampling from this proposal distribution may be difficult. Instead, the authors employ a Markov kernel targeting the GP posterior, taking advantage of the rich literature on MCMC methods for GP models.
\end{description} 
This is a large class of MH algorithms, many of which are not designed with their relation to IMH in mind. We mention that this excludes the Gaussian random walk proposal, which has no stationary probability distribution.

Our goal is to study convergence properties of this class of Metropolis--Hastings algorithms. In particular, we study the influence of two factors on the mixing properties of such algorithms: (i) the mixing properties of the proposal kernel with respect to its stationary distribution, and (ii) the mismatch between this stationary distribution and the target distribution of the overall MH kernel. 
Using a comparison of Dirichlet forms, we provide conditions for a Poincar\'e inequality for the proposal kernel (corresponding to the first ingredient) to be stable
under a Metropolis correction determined by the second ingredient.
We further discuss when this result on the right spectral gap yields $L^2$-geometric ergodicity of the Metropolis chain.

\subsection{Main results}\label{sec:intro-result}

Let $(\sX, \cB(\sX))$ be a measurable space, and let $\pi$ and $q$ be mutually absolutely continuous probability measures on $(\sX, \cB(\sX))$. Then there exists a measurable function $h:\sX \to \bR$ such that we may write
\begin{equation}
    \pi = \frac{1}{Z}e^h q,
\quad\text{with}\quad
Z=\int_{\sX}e^{h}\,dq. \label{eq:osc}
\end{equation}
Throughout the paper, equalities and inequalities between measurable functions on $\sX$ are understood to hold $q$-almost everywhere (equivalently, $\pi$-almost everywhere), while suprema and infima over measurable subsets of $\sX$ are interpreted as $q$-essential suprema and infima.

Consider a Markov kernel $K : \sX\times\cB(\sX)\rightarrow [0,1]$ that is $q$-reversible, so that
\begin{equation*}
    q(dx) K(x,dx') = q(dx') K(x',dx).
\end{equation*}
We will use $K$ as the proposal distribution in a Metropolis--Hastings update that targets $\pi$. For current state $x$ and proposed state $x'$, the acceptance probability $\alpha(x,x')$ equals
\begin{equation*}
    \alpha(x,x') = 1 \wedge \frac{\pi(dx') K(x',dx)}{\pi(dx) K(x,dx')}  =  1 \wedge \exp(h(x')-h(x)).
\end{equation*}
The second equality above follows from the two preceding equations. Note that this is also the acceptance probability of the independent MH algorithm where proposals are generated directly from $q$. Thus, even if the transition probability $K$ is quite involved, the acceptance probability depends only on its stationary distribution $q$.
Define the Metropolis Markov kernel
\begin{equation}
    P_K(x,dx') = K(x,dx')\alpha(x,x') + r(x)\delta_x(dx'), \label{def:mhkernel}
\end{equation}
where $r(x) := 1 - \int K(x,dy)\alpha(x,y)$,
and $\delta_x(\cdot)$ denotes the Dirac measure at $x$. In words, $P_K$ is a Metropolis kernel targeting $\pi$ via proposal kernel $K$ whose stationary distribution is $q$. Replacing $K(x,dx')$ with $q(dx')$, we obtain $P_q$, the independent Metropolis kernel with proposal $q$.

In~\Cref{sec:inheritance}, we investigate the inheritance of spectral gaps from the reference kernel $K$ to the Metropolis kernel $P_K$, and how this is affected by the relationship between the two stationary distributions $\pi$ and $q$. We first consider the case where the perturbation $h$ has bounded oscillation. Under this assumption, the Radon--Nikodym derivative $d\pi/dq$ is bounded above and away from zero, which allows us to compare the Dirichlet forms of $K$ and $P_K$ directly. A Holley--Stroock perturbation argument then yields explicit upper and lower bounds on the right spectral gap of $P_K$ in terms of the right spectral gap of $K$ and the oscillation of the perturbation. In particular, $P_K$ has a nonzero right spectral gap whenever $K$ does. This is a consequence of
\citet[Proposition~2.3]{madras99is}.

The problem becomes delicate when the oscillation of the perturbation is unbounded. A direct comparison of Dirichlet forms over the full state space yields no nontrivial spectral gap bound. The natural question is whether the inheritance of spectral gaps can still be established under verifiable conditions. Our main result (\Cref{thm:HS_ubdd}) relaxes the bounded oscillation assumption by allowing the perturbation to be unbounded from below while remaining bounded above. This is a common setting in Bayesian inference, where $q$ represents the prior distribution and $h$ represents the log likelihood. We provide sufficient conditions under which a Poincar\'e inequality for $K$ remains valid for $P_K$; equivalently, $P_K$ inherits a nonzero right spectral gap from $K$. At a high level, the proof of~\Cref{thm:HS_ubdd} involves a decomposition of the state space into a superlevel set of the perturbation and its complement. We then apply a Holley--Stroock perturbation argument on the superlevel set and combine it with a Lyapunov drift condition. For suitable superlevel sets, we control the transitions of $K$ across their boundaries and the rejection probability of $P_K$ outside these sets, which together ensure the stability of the Poincar\'e inequality. These results further imply the $L^2$-geometric convergence of $P_K$, either directly when $K$ is positive or after a lazy modification of $P_K$.

We also discuss the inheritance of uniform ergodicity, which, for reversible Markov kernels, implies a nonzero $L^2$-spectral gap. The question now is whether the uniform minorization condition of the reference kernel persists under the Metropolis correction. For the transfer of a one-step uniform minorization condition, an upper bound on the perturbation suffices. For a general multi-step uniform minorization condition, however, we require the perturbation to have bounded oscillation.

In~\Cref{sec:pcn}, we apply~\Cref{thm:HS_ubdd} to the pCN algorithm for sampling probability measures on a Hilbert space. Specifically,~\Cref{thm:dim_ind_gap_pcn} establishes a dimension-independent $L^2$-spectral gap for the pCN algorithm under the assumption that the potential function is locally bounded and \emph{quasiconvex}. Through an application to a Bayesian inverse problem with a Gaussian prior and Poisson data, we show that our criterion applies in a regime where the regularity conditions used in~\citet{hairer14gappcn} fail.

\subsection{Related literature}\label{sec:intro-literature}

Our work is closely related to~\citet{vollmer15invnongaussian}, which shows the inheritance of right spectral gap from the proposal kernel to the Metropolis kernel when the density of the target measure with respect to the reference measure is bounded above and away from zero.
However, the resulting lower and upper bounds on the spectral gap of the Metropolis kernel respectively involve the fourth power of the ratio of the essential lower to upper bounds on the density and the third power of its reciprocal. Using the result of \citet{madras99is}, we show that sharper bounds can be obtained which involve only the corresponding first powers; this order agrees with that obtained by~\cite{yali96gibbsgaussian} for Gibbs sampler targeting a perturbation of a Gaussian measure.

Our spectral gap inheritance results build on prior works establishing the stability of Poincar\'e-Sobolev type inequalities under perturbations. The first result of this kind was established in the celebrated work of~\citet{holley87lsi} for the logarithmic Sobolev inequality. For unbounded perturbations, the strategy of combining a local Poincar\'e inequality with a Lyapunov drift condition to establish a global inequality has been studied for certain classes of continuous-time Markov processes~\citep{bakry08poincare,bakry08ergodicdiffusion}. We note that \citet[Corollary 4.1]{bakry08ergodicdiffusion} use boundedness of the potential on a ball to obtain a local Poincar\'e inequality for the invariant measure of an overdamped Langevin diffusion by comparison with Lebesgue measure, that is similar in spirit to the localized Holley--Stroock argument used in this study. The Lyapunov criterion for establishing a Poincar\'e inequality was adapted to reversible Markov chains by \citet{taghvaei22lyapunovpoincare} and subsequently extended by \citet{andrieu22pi}, whose result we use in our analysis.

Several works have established spectral gaps for Metropolis–Hastings algorithms on infinite-dimensional spaces. \citet{hairer14gappcn} establish a Wasserstein spectral gap for the pCN algorithm for a class of target measures, with a bound uniform in the truncation level. In their setting, this also yields an $L^2$-spectral gap. \citet{hosseini23gapinfmh} show a spectral gap of a class of Metropolis–Hastings algorithms for non-Gaussian prior measures in a Wasserstein-like semimetric. Both~\citet{hairer14gappcn} and~\citet{hosseini23gapinfmh} apply the weak Harris theorem~\citep{hairer2011asymptotic}. Their analyses impose bounds on the growth rate of the potential's local Lipschitz constant and require a uniform lower bound on the acceptance probability for proposals in a ball around the current state of the chain. In this
study, we approach the problem from a different angle, inspecting whether the behavior of the potential in the tails is compatible with the mixing properties of the pCN proposal. This perspective leads to different conditions on the potential: quasiconvexity and local boundedness, rather than local Lipschitzness.

\section{Poincar\'e inequalities and Lyapunov conditions}\label{sec:preliminary}

In this section, we review the $L^2$ theory for Markov chains, and state the Lyapunov criterion for Poincar\'e inequality of a reversible Markov kernel developed by~\citet{taghvaei22lyapunovpoincare} and~\citet{andrieu22pi}.

Let $\nu$ be a probability measure on $(\sX,\cB(\sX))$ and $T: \sX\times \cB(\sX) \rightarrow [0,1]$ be a reversible Markov kernel with $\nu$ as its stationary measure, i.e., $\nu T(\cdot) := \int_{\sX} \nu(dx)T(x,\cdot) = \nu(\cdot)$. We state our results in this section generically for the pair $(\nu, T)$; in the rest of the paper, we will apply them to $(K,q)$, and $(P_K,\pi)$. The Hilbert space $L^2(\nu)$ consists of square-integrable functions with respect to $\nu$, i.e., $f : \sX \rightarrow \bR$ such that $\int_{\sX}f(x)^2\nu(dx) < \infty$. For $f,g \in L^2(\nu)$, the inner product is defined as $\langle f,g\rangle_{\nu} = \int_{\sX} f(x)g(x)\nu(dx)$, and the corresponding norm is given by $\Vert f\Vert_{\nu} = \sqrt{\langle f,f\rangle_{\nu}}$. Let $L_0^2(\nu)$ denote the subspace of $L^2(\nu)$ consisting of $\nu$-centered functions $f$, that is, $\nu f := \int_{\sX}f(x)\nu(dx) = 0$. We write $Tf(\cdot) = \int_{\sX} T(\cdot,dx')f(x')$ for the action of $T$ on $f \in L^2(\nu)$. We call $T$ self-adjoint when $\langle Tf,g\rangle_\nu = \langle f,Tg\rangle_\nu$ for $f,g \in L_0^2(\nu)$, which is equivalent to the reversibility of the kernel. We say $T$ is positive if it is self-adjoint, and $\langle f, Tf\rangle_\nu \geq 0$ for all $f\in L_0^2(\nu)$. The operator norm of $T$ is defined by
\begin{equation*}
    \Vert T \Vert_\nu := \sup_{f\in L^2_0(\nu)\backslash\{0\}} \frac{\Vert Tf\Vert_\nu}{\Vert f \Vert_\nu}.
\end{equation*}

Let $\sigma(T)$ denote the spectrum of $T$ with respect to $L^2_0(\nu)$. The self-adjointness of $T$ gives $\sigma(T) \subseteq [-1,1]$. Write $\lambda_{\max}(T)$ and $\lambda_{\min}(T)$ for the sharp upper and lower bound on the spectrum $\sigma(T)$, which are defined by
\begin{equation*}
    \lambda_{\max}(T) := \sup_{f\in L^2_0(\nu)\backslash\{0\}} \frac{\langle Tf, f\rangle_\nu}{\Vert f \Vert^2_\nu},\qquad \lambda_{\min}(T) :=  \inf_{f\in L^2_0(\nu)\backslash\{0\}} \frac{\langle Tf, f\rangle_\nu}{\Vert f \Vert^2_\nu}.
\end{equation*}
The right and left spectral gap are defined by $\gap_{\rR}(T) := 1 - \lambda_{\max}(T)$ and $\gap_{\rL}(T) := 1 + \lambda_{\min}(T)$. The (absolute) $L^2_\nu$-spectral gap of $T$, denoted by $\gap(T)$, is defined to be one minus the spectral radius of $T$, and is equal to the minimum of right and left spectral gaps. Since $T$ is self-adjoint, $\Vert T \Vert_\nu = |\lambda_{\max}(T)| \vee |\lambda_{\min}(T)|$, and hence $\gap(T) = 1 - \Vert T \Vert_\nu$.

For $f\in L^2(\nu)$, the Dirichlet form associated with the pair $(T,\nu)$ is defined by
\begin{equation}
    \cE_{T}(f,f) := \langle (\rI-T)f,f \rangle_\nu = \frac{1}{2}\iint (f(x)-f(x'))^2 \nu(dx)T(x,dx'), \label{eq:dirichlet_general}
\end{equation}
where $\rI$ denotes the identity operator in $L^2(\nu)$. Suppose that $T$ is a Metropolis kernel written as
\begin{equation*}
    T(x,dx') = Q(x,dx')a(x,x') + \left(1-\int Q(x,dy)a(x,y)\right)\delta_x(dx'),
\end{equation*}
where $Q : \sX\times\cB(\sX)\rightarrow [0,1]$ is the proposal kernel, and $a : \sX\times\sX\rightarrow [0,1]$ is the acceptance probability. Since a rejected proposal leaves the state of the chain unchanged at $x$, the rejection term does not contribute to the Dirichlet form. Hence, by~\eqref{eq:dirichlet_general},
\begin{equation}
    \cE_T(f,f) = \frac{1}{2} \iint (f(x)-f(x'))^2 \nu(dx)Q(x,dx')a(x,x').\label{eq:dirichlet_metropolis}
\end{equation}
We say $T$ satisfies a Poincar\'e inequality with constant $C \in (0,\infty)$ if
\begin{equation}
    C \cE_T(f,f) \geq\var_\nu(f)\qquad \forall f\in L^2(\nu), \label{eq:PI}
\end{equation}
where $\var_{\nu}(f) = \Vert f-\nu f\Vert^2_\nu$. The right spectral gap of $T$ admits the following representation
\begin{equation}
    \gap_{\rR}(T) = \underset{\substack{f\in L^2(\nu)\\ \text{s.t.}\var_\nu(f)>0}}{\inf}\frac{\cE_{T}(f,f)}{\var_{\nu}(f)}. \label{eq:rightgap}
\end{equation}
It follows from~\eqref{eq:PI} and~\eqref{eq:rightgap} that $T$ satisfies the Poincar\'e inequality if and only if $\gap_{\rR}(T) > 0$. In this case, the Poincar\'e constant, defined as the smallest admissible value of $C$ in~\eqref{eq:PI}, is $1/\gap_{\rR}(T)$.

To state the Lyapunov drift condition for establishing a Poincar\'e inequality, we first define the restriction of a Markov kernel. Let $A\in \cB(\sX)$ be such that $\nu(A) > 0$. The restriction of $T$ to $A$ is the Markov kernel on $A$ defined by 
\begin{equation}
    T|_A(x,dx') = T(x,dx')\ind_A(x') + \delta_x(dx')T(x,A^c),\quad x\in A. \label{def:restricted_chain}
\end{equation}
In words, the restricted kernel $T|_{A}(x,\cdot)$ is a Metropolis Markov kernel with proposal kernel $T$ and target measure $\nu|_A$, where $\nu|_A(B) := \nu(A \cap B)/\nu(A)$ for $B\in \cB(\sX)$. By construction, the restricted kernel $T|_A$ is $\nu|_A$-reversible.

\citet{taghvaei22lyapunovpoincare} developed a Lyapunov criterion for establishing a Poincar\'e inequality for reversible Markov chains. \citet{andrieu22pi} extended this approach by replacing the minorization condition with a local Poincar\'e inequality. We will use the following slight variant of their result, formulated in terms of a ratio drift condition that is more suitable for our purposes. For the reader’s convenience, we restate the result and its proof in the notation of our setting.

\begin{lemma}[A Lyapunov criterion and Poincar\'e inequality for reversible Markov kernels~\citep{taghvaei22lyapunovpoincare,andrieu22pi}]
    \label{lem:Lyapunov_Poincare}
    Suppose that there exist constants $\lambda \in (0,1]$, $b \in (0,\infty)$, a measurable set $U \in \cB(\sX)$, and a measurable function $V : \sX \rightarrow [1,\infty)$ such that
    \begin{equation}
        TV/V \leq (1-\lambda) + b\ind_U, \label{eq:ratio_drift}
    \end{equation}
    and that the restriction of $T$ to $U$ satisfies a Poincar\'e inequality: there exists $C_0 \in (0,\infty)$ such that
    \begin{equation}
        C_0 \cE_{T|_U}(f,f) \geq \var_{\nu|_U}(f)\qquad \forall f\in L^2(\nu|_U). \label{eq:PI_res_kernel}
    \end{equation}
    Then we have the following Poincar\'e inequality for $T$:
    \begin{equation*}
        C\cE_T(f,f) \geq \var_\nu(f) \qquad\forall f\in L^2(\nu),
    \end{equation*}
    with $C=(1+C_0b)/\lambda$.
\end{lemma}

\begin{proof}
    Fix $f\in L^2(\nu)$ and set $f_0 = f - \nu|_U(f)$. The Poincar\'e inequality for $T|_U$ in~\eqref{eq:PI_res_kernel}, together with Lemma 69 of~\citet{andrieu22pi}, gives
    \begin{equation}
        \langle f_0^2,\ind_U\rangle_\nu \leq C_0\cE_T(f,f). \label{eq:PI_local}
    \end{equation}
    The drift condition in~\eqref{eq:ratio_drift} implies
    \begin{equation}
        \lambda \leq 1-\frac{TV}{V} + b\ind_U,\quad\text{and hence}\quad f_0^2 \leq \frac{f_0^2}{\lambda}\left(1-\frac{TV}{V}\right) + \frac{b}{\lambda}f_0^2\ind_U. \label{eq:andrieuthm70}
    \end{equation}
    Moreover, the drift condition gives $\sup_{x\in \sX}TV/V(x) \leq 1+b < \infty$. Since $f_0$ differs from $f$ by a constant, $\cE_T(f_0,f_0) = \cE_T(f,f)$. Lemma 71 of~\citet{andrieu22pi} therefore yields
    \begin{equation}
        \langle f_0^2,1-TV/V\rangle_\nu \leq \cE_T(f,f). \label{eq:andrieulem71}
    \end{equation}
    Finally, by the variational characterization $\nu f = \argmin_{c\in \bR}\Vert f-c\Vert^2_\nu$,
    \begin{align*}
        \Vert f -\nu f\Vert^2_\nu &\leq \Vert f_0 \Vert_\nu^2 \quad\text{(since~$f_0 = f - \nu|_U(f)$)}\\
        &\leq \frac{1}{\lambda}\langle f_0^2,1-TV/V\rangle_\nu + \frac{b}{\lambda}\langle f_0^2,\ind_U\rangle_\nu\quad\text{(by~\eqref{eq:andrieuthm70})}\\
        &\leq \frac{1+C_0b}{\lambda}\cE_T(f,f)\quad\text{(by~\eqref{eq:PI_local} and~\eqref{eq:andrieulem71})},
    \end{align*}
    which completes the proof.
\end{proof}

\section{Inheritance of the spectral gaps}\label{sec:inheritance}

In this section, we present our main theoretical results on the transference of spectral gaps from the reference kernel $K$ to the Metropolis kernel $P_K$ defined in~\eqref{def:mhkernel}. In~\Cref{sec:inheritance-PI}, we state the assumptions and results concerning the inheritance of the right spectral gap. In~\Cref{sec:inheritance-abs}, we explain how these results yield a nonzero $L^2$-spectral gap for $P_K$, either through a slight modification of the chain or directly when $K$ is positive.

\subsection{Stability of the Poincar\'e inequality}\label{sec:inheritance-PI}

First, we study the inheritance of the right spectral gap from the reference kernel $K$ to the Metropolis kernel $P_K$. Define the oscillation of the perturbation $h$ from~\eqref{eq:osc} as
\begin{equation*}
    \osc(h) := M_h - m_h\quad\text{where}\quad M_h := \sup_{x\in \sX} h,\quad\text{and}\quad m_h := \inf_{x\in \sX} h.
\end{equation*}
We begin with the bounded oscillation case, i.e., $\osc(h) < \infty$, and then extend the analysis to an unbounded oscillation case.

In the bounded oscillation case, the perturbation induces only a uniformly bounded change of measure. Consequently, the spectral gap of the Metropolis kernel differs from that of the reference kernel by at most a multiplicative factor $e^{\osc(h)}$. This is a consequence of \citet[Proposition 2.3]{madras99is}, which compares the spectral gaps of two Metropolis chains sharing a common symmetric proposal kernel. Our setting is slightly different, and for completeness, we provide a self-contained proof below:
\begin{theorem}[Holley--Stroock]
    \label{thm:HS}
    Assume $\osc(h) < \infty$ and $\gap_{\rR}(K) > 0$. Then
    \begin{equation*}
        e^{-\osc(h)}\gap_{\rR}(K) \leq \gap_{\rR}(P_K) \leq e^{\osc(h)}\gap_{\rR}(K).
    \end{equation*}
\end{theorem}

\begin{proof}
    Since $h$ has bounded oscillation, we have
    \begin{equation*}
        0 < \frac{e^{m_h}}{Z} \leq \frac{d\pi}{dq} \leq \frac{e^{M_h}}{Z} < \infty,\quad\text{and}\quad L^2(\pi) = L^2(q).
    \end{equation*}
    
    Fix $f \in L^2(\pi)$. From~\eqref{eq:dirichlet_metropolis}, the Dirichlet form of $P_K$ satisfies
    \begin{align}
        2\cE_{P_K}(f,f) &= \iint \pi(dx)K(x,dx')[1\wedge e^{h(x')-h(x)}](f(x')-f(x))^2 \nonumber\\
        &= \frac{1}{Z}\iint q(dx)K(x,dx')[e^{h(x)}\wedge e^{h(x')}](f(x')-f(x))^2 \nonumber\\
        &\geq \frac{e^{m_h}}{Z}\iint q(dx)K(x,dx')(f(x')-f(x))^2 \nonumber\\
        &= 2\frac{e^{m_h}}{Z}\cE_K(f,f). \label{eq:HS_dir_bound} 
    \end{align}
    
    By assumption, $K$ satisfies a Poincar\'e inequality, i.e., 
    \begin{equation}
        \cE_K(f,f) \geq \gap_{\rR}(K) \var_q(f). \label{eq:HS_PI_K}
    \end{equation}
    
    Using the characterization $\var_\pi(f) = \inf_{c\in\bR}\Vert f-c \Vert_\pi^2$ and the bound $d\pi/dq \leq e^{M_h}/Z$, we obtain
    \begin{equation}
        \var_\pi(f) \leq \Vert f-qf\Vert^2_\pi = \int (f - qf)^2\frac{e^{h}}{Z}dq \leq \frac{e^{M_h}}{Z}\var_q(f). \label{eq:HS_var_bound}
    \end{equation}
    Combining~\eqref{eq:HS_dir_bound},\eqref{eq:HS_PI_K}, and \eqref{eq:HS_var_bound} yields
    \begin{align*}
        \cE_{P_K}(f,f) &\geq \frac{e^{m_h}}{Z} \gap_{\rR}(K) \frac{Z}{e^{M_h}}\var_{\pi}(f) 
        = e^{-\osc(h)}\gap_{\rR}(K)\var_\pi(f),
    \end{align*}
    and thus
    \begin{equation*}
        \gap_{\rR}(P_K) \geq e^{-\osc(h)}\gap_{\rR}(K).
    \end{equation*}
    A similar argument gives
    \begin{equation*}
        \cE_{P_K}(f,f)\le \frac{e^{M_h}}{Z}\cE_K(f,f) \quad\text{and}\quad \var_{\pi}(f)\ge \frac{e^{m_h}}{Z}\var_q(f),
    \end{equation*}
    and thus
    \begin{equation*}
        \gap_{\rR}(P_K) \le e^{M_h -m_h}\gap_{\rR}(K) =e^{\osc(h)}\gap_{\rR}(K),
    \end{equation*}
    which completes the proof.
\end{proof}

\begin{remark}
    The result of~\Cref{thm:HS} implies that $\gap_{\rR}(P_K) > 0$. Hence $P_K$ also satisfies a Poincar\'e inequality.
\end{remark}

We now turn our attention to the unbounded oscillation case. In particular, we assume that $M_h < \infty$ while $m_h = -\infty$. In this case, $\osc(h) = \infty$, so the comparison bound of~\Cref{thm:HS} becomes vacuous. Nevertheless, under the assumptions introduced below, $P_K$ can still be shown to satisfy a Poincar\'e inequality.

We begin by introducing some notation. For each $\tau > 0$, define the superlevel set of the perturbation
\begin{equation*}
    U_{\tau} := \{ x\in \sX : h(x) > -\tau\}.
\end{equation*}
Let $q_\tau$ and $\pi_\tau$ denote the restriction of $q$ and $\pi$ to $U_{\tau}$, respectively. Similarly, we write $K_\tau$ and $P_{K,\tau}$ for the restriction of Markov kernels $K$ and $P_K$ to $U_{\tau}$, respectively.

We state the assumptions for~\Cref{thm:HS_ubdd}, which is the main result of this section. We first assume that the reference kernel $K$ satisfies the following drift condition.

\begin{assumption}
    \label{ass:uosc1}
    There exists a measurable function $V_0 : \sX\rightarrow [1,\infty)$ and a constant $\tau_0 > 0$ such that
    \begin{equation*}
        \underset{x\in U_{\tau_0}^c}{\sup}\frac{KV_0(x)}{V_0(x)} < 1\quad\text{and}\quad\underset{x\in \sX}{\sup}\frac{KV_0(x)}{V_0(x)} < \infty.
    \end{equation*}
\end{assumption}

\begin{remark}
    A simple sufficient condition for~\Cref{ass:uosc1} is that there exist constants $\lambda_0 \in (0,1]$, $b_0 < \infty$, and $\tau_0 > 0$ such that $KV_0 \leq (1-\lambda_0) V_0 + b_0\ind_{U_{\tau_0}}$. More generally, it suffices that $K$ satisfy the drift condition $KV_0 \leq (1-\lambda_0)V_0 + b_0\ind_C$ for some measurable set $C \subseteq U_{\tau_0}$, which need not be petite.
\end{remark}
 
The next assumption concerns the escape probability from $U_\tau$ under the kernel $K$.

\begin{assumption}
    \label{ass:uosc2}
    The escape probability from $U_\tau$ under $K$ vanishes uniformly over $U_\tau$ as $\tau \rightarrow \infty$; that is,
    \begin{equation*}
        \underset{\tau\rightarrow\infty}{\lim}\underset{x\in U_{\tau}}{\sup}K(x,U_{\tau}^{c}) = 0.
    \end{equation*}
\end{assumption}

In other words, the probability that a proposal started from $x \in U_\tau$ leaves $U_\tau$ tends to zero uniformly over $U_\tau$ as $\tau \rightarrow \infty$. This condition holds for any independent proposal $K$ (where $K(x,\cdot) = q(\cdot)$). We can also expect it to hold for Markov kernels targeting $q$ so long as the tails of $q$ are appropriately aligned with $h$. Finally, we impose a tightness assumption on the rejection probability of Metropolis kernel $P_K$.

\begin{assumption}
    \label{ass:uosc3}
    The rejection probability under Metropolis kernel $P_K$ vanishes uniformly over $U_\tau^c$ as $\tau \rightarrow \infty$; that is,
    \begin{equation*}
        \underset{\tau\rightarrow\infty}{\lim}\underset{x\in U_\tau^c}{\sup} \int K(x,dx')(1-\alpha(x,x')) = 0.
    \end{equation*}
\end{assumption}
Hence, the Metropolis kernel remains uniformly close to the reference kernel in tail $U_\tau^c$. This ensures that rejections induced by the Metropolis correction do not weaken the drift inherited from $K$ on $U_\tau^c$. 

{
\begin{remark}
    \Cref{ass:uosc3} is a slightly stronger condition than required. 
    The proof of~\Cref{lem:driftinherit} just requires that, for all sufficiently large $\tau$, $\sup_{x\in U_\tau^c} \int K(x,dx')(1-\alpha(x,x')) < \lambda_0$ where $\lambda_0 := 1 - \sup_{x\in U^c_{\tau_0}} KV_0(x)/V_0(x) > 0$ is given by~\Cref{ass:uosc1}. We impose this stronger condition as it is slightly easier to verify.
\end{remark}
}

\begin{theorem}
    \label{thm:HS_ubdd}
    Under~\Cref{ass:uosc1,ass:uosc2,ass:uosc3} and $M_h < \infty$, if $K$ satisfies a Poincar\'e inequality, then so does $P_K$.
\end{theorem}

We first introduce the key lemmas that will be used to prove~\Cref{thm:HS_ubdd}. \Cref{lem:resPI} establishes a Poincar\'e inequality for $K_\tau$. The key observation is that, for sufficiently large $\tau$, the transitions removed by the restriction to $U_\tau$ contribute only negligibly to the Dirichlet form. Consequently, a Poincar\'e inequality for $K$ transfers to $K_\tau$. \citet[Lemma 18]{rudolf_18_gpCN} derive an analogous result for the $L^2$-spectral gap. We adapt their zero extension argument to our setting.

\begin{lemma}[Spectral gap of restricted chain]
    \label{lem:resPI}
    Suppose that $\gap_{\rR}(K) > 0$ and~\Cref{ass:uosc2} holds. Then there exists $\tau_1 > 0$ such that, for every $\tau \geq \tau_1$, $K_\tau$ satisfies a Poincar\'e inequality with 
    \begin{equation*}
        \gap_{\rR}(K_\tau) \geq \gap_{\rR}(K) - \underset{x\in U_{\tau}}{\sup}K(x,U_{\tau}^{c}) > 0.
    \end{equation*}
\end{lemma}

\begin{proof}
    Fix $\tau > 0$ and $f\in L_0^2(q_\tau)$; recall that $q_\tau$ is the restriction of $q$ to $U_\tau$. Define $\ff := f\ind_{U_{\tau}} \in L^2_0(q)$. Since $K$ satisfies a Poincar\'e inequality, we have
    \begin{equation}
        C_K\cE_K(\ff,\ff) \geq \var_{q}(\ff), \label{eq:PI_unres_K}
    \end{equation}
    with $C_K := 1/\gap_{\rR}(K) > 0$.

    We decompose the double integral in the Dirichlet form into the four nonoverlapping regions $U_\tau \times U_\tau, U_\tau \times U_\tau^c, U_\tau^c \times U_{\tau}$ and $U_\tau^c \times U_\tau^c$. Thus, with $I_{A,B} := \iint_{A\times B} q(dx)K(x,dx')(\ff(x')-\ff(x))^2,\quad A,B \in \cB(\sX)$, we have
    \begin{align*}
        \cE_K(\ff,\ff) &= \frac{1}{2}(I_{U_\tau,U_\tau} + I_{U_\tau,U_\tau^c} + I_{U_\tau^c,U_\tau} + I_{U_\tau^c,U_\tau^c}).
    \end{align*} 
    Observe that on $U_\tau$, $\ff = f$, and on $U_\tau^c$, $\ff = 0$. Thus,
    \begin{align*}
        I_{U_\tau,U_\tau} &= q(U_{\tau})\iint q_\tau(dx)K_\tau(x,dx')(f(x')-f(x))^2 = 2q(U_{\tau})\cE_{K_\tau}(f,f)\\
        I_{U_\tau,U_\tau^c} &= \iint_{U_\tau\times U_\tau^c} q(dx)K(x,dx')f^2(x)= \int_{U_\tau}q(dx)K(x,U_\tau^c)f^2(x).
    \end{align*}
    By reversibility of $K$, $I_{U_\tau,U_\tau^c} = I_{U_\tau^c,U_\tau}$. Combining these identities yields
    \begin{align}
        \cE_K(\ff,\ff) &= q(U_\tau)\cE_{K_\tau}(f,f) + \int_{U_\tau} q(dx)K(x,U^c_\tau)f^2(x) \nonumber\\
        &\leq q(U_\tau)\cE_{K_\tau}(f,f) + q(U_\tau)\varepsilon_\tau q_\tau(f^2) \quad (\text{with }     \varepsilon_\tau := \sup_{x\in U_\tau}K(x,U^c_\tau) )\nonumber\\
        &= q(U_\tau)\left\{\cE_{K_\tau}(f,f) + \varepsilon_\tau\var_{q_\tau}(f)\right\} \quad (\text{since } f\in L_0^2(q_\tau)). \label{eq:res_dir_bdd}
    \end{align}
    
    We also have
    \begin{align}
        \var_q(\ff) &= q(f^2\ind_{U_\tau}) - q(f\ind_{U_\tau})^2\nonumber\\
        &= q(U_{\tau})q_\tau(f^2) - q(U_{\tau})^2q_\tau(f)^2\nonumber\\
        &= q(U_{\tau})\var_{q_\tau}(f). \label{eq:res_var_bdd}
    \end{align}
    Substituting~\eqref{eq:res_dir_bdd} and \eqref{eq:res_var_bdd} to \eqref{eq:PI_unres_K} and dividing both sides by $q(U_\tau)$, we have
    \begin{equation*}
        \var_{q_\tau}(f) \leq C_K\cE_{K_\tau}(f,f) + C_K\varepsilon_\tau\var_{q_\tau}(f).
    \end{equation*}
    By~\Cref{ass:uosc2}, there exists a $\tau_1 > 0$ such that for every $\tau \geq \tau_1$, we have $\varepsilon_{\tau} < 1/C_K$, and so
    \begin{equation*}
        \cE_{K_\tau}(f,f) \geq \frac{1-C_K\varepsilon_\tau}{C_K}\var_{q_\tau}(f),
    \end{equation*}
    and
    \begin{equation*}        
        \gap_{\rR}(K_\tau) \geq \frac{1-C_K\varepsilon_\tau}{C_K} = \gap_{\rR}(K) - \varepsilon_\tau > 0.
    \end{equation*}
    We thus have the desired result.
\end{proof}

In fact, the above proof implies the following stronger conclusion:
\begin{corollary}
Under the assumptions of \Cref{lem:resPI},
$ \liminf_{\tau\to\infty}
    \gap_{\rR}(K_\tau)
    \geq
    \gap_{\rR}(K)$.
\end{corollary}
Next, we present~\Cref{lem:driftinherit}, which establishes the inheritance of the drift condition from $K$ to $P_K$. The intuition is that, for a sufficiently large $\tau$, if the reference kernel $K$ drifts toward the localization set $U_\tau$ and the acceptance probability of the Metropolis kernel $P_K$ approaches unity on the exterior region $U_\tau^c$, then $P_K$ inherits the drift condition of $K$.

\begin{lemma}[Inheritance of the drift condition]
    \label{lem:driftinherit}
    Under~\Cref{ass:uosc1,ass:uosc3}, there exists a measurable function $V : \sX\rightarrow [1,\infty)$ and a constant $\tau_2 > 0$ such that, for every $\tau \geq \tau_2$,
    \begin{equation*}
        \underset{x\in U_\tau^c}{\sup}\frac{P_KV(x)}{V(x)} < 1\quad\text{and}\quad\underset{x\in \sX}{\sup}\frac{P_KV(x)}{V(x)} < \infty.
    \end{equation*}
\end{lemma}

\begin{proof}
    By~\Cref{ass:uosc1}, there exists a measurable function $V_0 : \sX\rightarrow [1,\infty)$ and constants $\tau_0 > 0$, $\lambda_0 \in (0,1]$, and $b_0 < \infty$ such that 
    \begin{equation*}
        \lambda_0 = 1-\underset{x\in U_{\tau_0}^c}{\sup}\frac{KV_0(x)}{V_0(x)} \quad\text{and}\quad b_0 = \underset{x\in \sX}{\sup}\frac{KV_0(x)}{V_0(x)}.
    \end{equation*}
    
    Set $V := V_0$. By the definition of $P_K$, we have
    \begin{align*} 
        \frac{P_KV(x)}{V(x)} &= \int K(x,dx')\alpha(x,x')\frac{V(x')}{V(x)} + r(x)\\
        &\leq \int K(x,dx')\frac{V(x')}{V(x)} + r(x) = \frac{KV(x)}{V(x)} + r(x).
    \end{align*}
    
    By~\Cref{ass:uosc3}, we may choose a $\tau_2 \geq \tau_0$ such that
    \begin{align*}
        r_{\tau_2} := \sup_{x\in U_{\tau_2}^c} r(x) = \underset{x\in U_{\tau_2}^c}{\sup} \int K(x,dx')(1-\alpha(x,x')) < \lambda_0.
    \end{align*}
    
    For $x\in U_{\tau_2}^c \subseteq U^c_{\tau_0}$, we have
    \begin{equation}
        \frac{P_KV(x)}{V(x)} \leq \frac{KV(x)}{V(x)} + r(x) \leq 1-\lambda_0 + r_{\tau_2} < 1. \label{eq:lyapunov_contraction}
    \end{equation}
    On the other hand, we have
    \begin{equation} 
        \frac{P_KV(x)}{V(x)} \leq \frac{KV(x)}{V(x)} + r(x) \leq b_0 + 1 < \infty\qquad \forall x \in \sX. \label{eq:lyapunov_bdd}
    \end{equation}
    Combining the two estimates in~\eqref{eq:lyapunov_contraction} and~\eqref{eq:lyapunov_bdd} completes the proof.
\end{proof}

\begin{remark}
    If we can assume that $V$ is bounded on $U_\tau$ for every $\tau > 0$, one can obtain the additive form of drift condition $P_KV \leq (1-\lambda) V + b\ind_{U_\tau}$.
\end{remark}

Now, we prove the~\Cref{thm:HS_ubdd}. The proof follows the same strategy as that of~\Cref{thm:HS}. We, however, first restrict the kernels to $U_\tau$ and apply the Holley--Stroock argument on the localized state space. The resulting local Poincar\'e inequality is then lifted to the full state space via the drift condition.

\begin{proof}[Proof of~\Cref{thm:HS_ubdd}]
    Let $\tau_1$ and $\tau_2$ be as in~\Cref{lem:resPI,lem:driftinherit}, respectively, and fix $\tau > 0$ such that $\tau \geq \tau_1 \vee \tau_2$ and $\pi(U_\tau) > 0$.  Recall that $P_{K,\tau}$ is the restriction of $P_K$ to $U_\tau$, which admits a two-stage Metropolis algorithm interpretation: the first stage targets $\pi$ using $K$ as the proposal kernel, and the second stage targets $\pi_\tau$ using $P_K$ as the proposal kernel. Using~\eqref{def:mhkernel} and~\eqref{def:restricted_chain}, for $x\in U_\tau$, we obtain
    \begin{align*}
        P_{K,\tau}(x,dx') &= P_K(x,dx')\ind_{U_\tau}(x') + \delta_{x}(dx')P_K(x,U_\tau^c)\\
        &= K(x,dx')[1 \wedge e^{h(x')-h(x)}]\ind_{U_\tau}(x') + \delta_x(dx')\underbrace{\{r(x) + P_K(x,U_\tau^c)\}}_{=:\zeta_\tau(x)}.
    \end{align*}
    Here, we used $\delta_x(dx')\ind_{U_\tau}(x') = \delta_x(dx')$ for $x\in U_\tau$ in the second equation. As before, the rejection term $\delta_x(dx')\zeta_\tau(x)$ makes no contribution to the Dirichlet form. Consequently, for $f\in L^2_0(\pi_\tau)$,
    \begin{align}
        2\cE_{P_{K,\tau}}(f,f) &=\iint \pi_\tau(dx)P_{K,\tau}(x,dx')(f(x')-f(x))^2\nonumber\\
        &= \iint \pi_\tau(dx)K(x,dx')[1 \wedge e^{h(x')-h(x)}]\ind_{U_\tau}(x')(f(x')-f(x))^2\nonumber\\ 
        &= \iint \frac{\pi(dx)\ind_{U_\tau}(x)}{\pi(U_\tau)}K(x,dx')[1 \wedge e^{h(x')-h(x)}]\ind_{U_\tau}(x')(f(x')-f(x))^2 \nonumber\\ 
        &= \frac{1}{\pi(U_\tau)Z}\iint q(dx)\ind_{U_\tau}(x)K(x,dx')[e^{h(x')} \wedge e^{h(x)}]\ind_{U_\tau}(x')(f(x')-f(x))^2 \nonumber\\ 
        &\geq \frac{e^{-\tau}}{\pi(U_\tau)Z}\iint q(dx)\ind_{U_\tau}(x)K(x,dx')\ind_{U_\tau}(x')(f(x')-f(x))^2 \nonumber\\ 
        &=\frac{e^{-\tau}}{Z}\frac{q(U_\tau)}{\pi(U_\tau)}\iint q_\tau(dx)K_\tau(x,dx')(f(x')-f(x))^2 \nonumber\\
        &= 2\frac{e^{-\tau}}{Z}\frac{q(U_\tau)}{\pi(U_\tau)} \cE_{K_\tau}(f,f). \label{eq:HS_ubdd_dir_bound}
    \end{align}
    By~\Cref{lem:resPI}, we have the following Poincar\'e inequality for the restricted kernel $K_\tau$
    \begin{equation}
        C_{K,\tau}\cE_{K_\tau}(f,f) \geq \var_{q_\tau}(f), \label{eq:HS_ubdd_PI_K}
    \end{equation}
    for some constant $C_{K,\tau} > 0$. We also note that, for $x \in U_\tau$,
    \begin{equation*}
        \frac{d\pi_\tau}{dq_\tau}(x) = \frac{q(U_\tau)}{\pi(U_\tau)}\frac{d\pi}{dq}(x) = \frac{q(U_\tau)}{\pi(U_\tau)} \frac{e^{h(x)}}{Z}\leq \frac{q(U_\tau)}{\pi(U_\tau)} \frac{e^{M_h}}{Z},
    \end{equation*}
    and thus, similar to the argument in~\Cref{thm:HS},
    \begin{equation}
        \var_{\pi_\tau}(f) \leq \frac{q(U_\tau)}{\pi(U_\tau)}\frac{e^{M_h}}{Z}\var_{q_\tau}(f). \label{eq:HS_ubdd_var_bound}
    \end{equation}    
    Combining the preceding bounds~\eqref{eq:HS_ubdd_dir_bound},\eqref{eq:HS_ubdd_PI_K}, and \eqref{eq:HS_ubdd_var_bound}, we have
    \begin{equation}
        e^{M_h+\tau}C_{K,\tau}\cE_{P_{K,\tau}}(f,f) \geq \var_{\pi_\tau}(f),\label{eq:HS_ubdd_PI_resMH}
    \end{equation}
    and thus the restricted kernel $P_{K,\tau}$ also satisfies a Poincar\'e inequality. By~\Cref{lem:driftinherit}, there exists a measurable function $V : \sX \rightarrow [1,\infty)$, and constants $b < \infty$ and $\lambda \in (0,1]$ such that
    \begin{equation*}
        P_KV/V \leq (1-\lambda) + b\ind_{U_\tau}.
    \end{equation*}
    Finally, applying~\Cref{lem:Lyapunov_Poincare} yields the following Poincar\'e inequality for $P_K$:
    \begin{equation*}
        \widetilde{C}\cE_{P_K}(\ff,\ff)\geq \var_\pi(\ff),\qquad \forall \ff \in L^2(\pi),
    \end{equation*}
    with $\widetilde{C} = (1+be^{M_h+\tau}C_{K,\tau})/\lambda$.
\end{proof}

\subsection{Inheritance of the $L^2$-spectral gap}\label{sec:inheritance-abs}

\Cref{thm:HS,thm:HS_ubdd} establish a nonzero right spectral gap for $P_K$, which is equivalent to its variance bounding property (uniformly bounding the asymptotic variance of MCMC estimates by a constant multiple of their stationary variance), as well as the Markov chain central limit theorem (CLT) for all $L^2$ functionals of reversible chains~\citep[Theorem 7 and 14]{roberts08variancebounding}. However, this does not in general imply a nonzero $L^2$-spectral gap, since the spectrum may still approach $-1$. Extending the result in~\Cref{sec:inheritance-PI} to $L^2$-spectral gap will establish the $L^2$-geometric ergodicity for reversible chains~\citep[Theorem 2.1]{roberts97ge}, a key property in the convergence analysis of MCMC algorithms. However, such an extension would require tools beyond the Lyapunov--Poincar\'e criterion of~\citet{andrieu22pi} used here, and we leave this question for future work. We instead explain when a nonzero $L^2$-spectral gap for $P_K$ follows from its variance boundedness.

For a positive reversible Markov kernel, variance bounding is equivalent to $L^2$-geometric ergodicity~\citep{roberts08variancebounding}. Recall that a Markov operator $T : L^2(\nu) \rightarrow L^2(\nu)$ is positive if it is self-adjoint and $\langle Tf,f\rangle_\nu \geq 0$ for all $f \in L^2_0(\nu)$. If a reversible Markov kernel is positive, its spectrum is contained in $[0,1]$, and thus the $L^2$-spectral gap and right spectral gap coincide. \citet[Lemma 3.1]{rudolf13positivity} establish the inheritance of positivity from the proposal kernel to the Metropolis kernel on Euclidean state spaces, and their argument extends directly to our general state-space setting.

\begin{lemma}[Positivity of Metropolis kernels~\citep{rudolf13positivity}]
    \label{lem:inherit_positivity}
    If the reference kernel $K$ is positive on $L^2_0(q)$, the Metropolis kernel $P_K$ is positive on $L^2_0(\pi)$.
\end{lemma}

Combining~\Cref{lem:inherit_positivity} with the results in~\Cref{thm:HS,thm:HS_ubdd} yields the following corollary.

\begin{corollary}
    \label{coro:abs_gap_positivity}
    Under the assumptions of~\Cref{thm:HS} or~\Cref{thm:HS_ubdd}, and if $K$ is positive, the Metropolis kernel $P_K$ has a nonzero $L^2$-spectral gap.
\end{corollary}

In general, a standard way to address this issue is to consider a lazy modification of the chain. That is, at each step, the chain remains at its current state with probability $\varepsilon\in(0,1)$ and otherwise makes a transition according to the original kernel. This modification shifts the spectrum to the right, away from minus one, so that a nonzero right spectral gap implies a nonzero $L^2$-spectral gap.

\begin{remark}[{\citet[Theorem~2]{roberts08variancebounding}}]
    Suppose that $P_K$ is variance bounding. Then, for any $\varepsilon\in(0,1)$, the lazy version of the Metropolis kernel $P_K^{\mathrm{lazy},\varepsilon} := \varepsilon \rI + (1-\varepsilon)P_K$ is $L^2$-geometrically ergodic.
\end{remark}

Thus, $L^2$-geometric ergodicity can always be obtained from~\Cref{thm:HS,thm:HS_ubdd} after an arbitrarily small lazy modification on the chain.

\subsection{Inheritance of uniform ergodicity} \label{sec:inheritance-uniform}

We next make the stronger assumption that the proposal kernel $K$ is uniformly ergodic. Now, we can directly work with an equivalent uniform minorization condition~\citep[Theorem 16.0.1]{meyn2012markov}. Investigating the preservation of uniform minorization conditions under the Metropolis correction provides a more direct route for transferring spectral gaps from the proposal kernel to the Metropolis kernel.

We say that a Markov kernel $P$ with state space $\sX$ satisfies a uniform minorization condition (which is also known as Doeblin's condition), if there exists $m\in \bN$, $\varepsilon > 0$, and a probability measure $\nu$ on $\sX$ such that
\begin{equation}
    P^m(x,\cdot) \geq \varepsilon\nu(\cdot)\qquad \forall x \in \sX.\label{eq:doeblin}
\end{equation}

\begin{theorem}
    \label{thm:inherit_onestep_minor}
    If $K$ satisfies the uniform minorization condition in~\eqref{eq:doeblin} with $m=1$ and $M_h < \infty$, the $P_K$ also satisfies a one-step uniform minorization condition.
\end{theorem}

\begin{proof}
    By assumption, we have
    \begin{equation*}
        \alpha(x,y) = 1 \wedge \exp(h(y)-h(x)) \geq \exp(h(y)-M_h)\qquad \forall x,y \in \sX.
    \end{equation*}
    
    Therefore, for given $x\in \sX$ and measurable $A \subseteq \sX$,
    \begin{align*}
        P_K(x,A) &= \int_A K(x,dx')\alpha(x,x') + r(x)\delta_x(A)\\
        &\geq \frac{\varepsilon}{\exp(M_h)}\int_A \exp(h(x'))\nu(dx') = \frac{\varepsilon C_h}{\exp(M_h)}\nu_h(A),
    \end{align*}
    where
    $C_h := \int_{\sX} \exp(h(y))\nu(dy),\ \nu_h(A) := \frac{1}{C_h} \int_A \exp(h(y))\nu(dy)$. This completes the proof.
\end{proof}

The one-step case needs only an upper bound on the starting point ($M_h$) and absorbs the destination into the measure. The multi-step case has intermediate states that are simultaneously destinations (need to be reached) and starting points (need to be left), and when $h$ is unbounded below, both roles cannot be controlled simultaneously without additional assumptions. The following elementary lemma bounds this loss solely in terms of the oscillation of the perturbation and yields a sharper minorization bound than a naive stepwise estimate.
\begin{lemma}[Path oscillation bound]
    \label{lem:pathbound}
    Let $\{a_i\}_{i=0}^{n}$ be a sequence with $\underline{a} \leq a_i \leq \overline{a}$ for $0\leq i\leq n$. Then, with $a_+ := \max(a,0)$ for $a\in \bR$,
    \begin{equation*}
        \sum_{i=0}^{n-1}(a_{i+1}-a_{i})_+ \leq \lceil n/2 \rceil (\overline{a}-\underline{a}).
    \end{equation*}
\end{lemma}

\begin{proof}
    Let $I_+ := \{i \in \{0,...,n-1\} : a_{i+1} > a_i\}$. Consecutive indices in $I_+$ can be merged by noting that $(a_{i+2} - a_{i+1})_+ + (a_{i+1}-a_i)_+ = (a_{i+2}-a_{i})_+$ for $i$, $i+1 \in I_+$. Repeating this procedure, the positive increments are grouped into disjoint maximal runs. Each run contributes at most $\overline{a}-\underline{a}$, because the sum over a run telescopes to the difference between its endpoints. Since distinct maximal increasing runs must be separated by at least one non-increasing increment, there can be at most $\lceil n/2 \rceil$ such runs. Therefore
    \begin{equation*}
        \sum_{i=0}^{n-1} (a_{i+1}-a_{i})_+ = \sum_{i \in I_+}(a_{i+1}-a_{i}) \leq \lceil n/2 \rceil (\overline{a}-\underline{a}).
    \end{equation*}
\end{proof}

We now prove the inheritance of general multi-step uniform minorization.

\begin{theorem}
    \label{thm:inherit_multistep_minor}
    Assume that $K$ satisfies the uniform minorization condition in~\eqref{eq:doeblin} and $\osc(h) < \infty$. Then $P_K$ also satisfies a $m$-step uniform minorization condition.
\end{theorem}

\begin{proof}
    Fix $x_0 \in \sX$ and measurable $A \subseteq \sX$. Restricting to trajectories for which all $m$ proposed moves are accepted, we obtain
    \begin{equation}
        P^m_K(x_0,A) \geq \int_{\sX^{m-1}\times A} \prod_{j=0}^{m-1} K(x_j,dx_{j+1})\alpha(x_j,x_{j+1}).\label{eq:acc_mstep_kernel}
    \end{equation}
    For the first $m-1$ acceptance probabilities,
    \begin{align*}
        \prod_{j=0}^{m-2} \alpha(x_j,x_{j+1}) &= \exp\left\{-\sum_{j=0}^{m-2}(h(x_j)-h(x_{j+1}))_+\right\}\\
        &\geq \exp\left\{-\bigg\lceil \frac{m-1}{2} \bigg\rceil\osc(h)\right\},
    \end{align*}
    where the inequality follows from~\Cref{lem:pathbound}. As in the proof of~\Cref{thm:inherit_onestep_minor}, 
    \begin{equation*}
        \alpha(x_{m-1},x_m) \geq \exp(h(x_m) - M_h).
    \end{equation*}
    Combining these bounds in~\eqref{eq:acc_mstep_kernel} gives
    \begin{align*}
        P^m_K(x_0,A) &\geq \exp(-\lceil (m-1)/2 \rceil \osc(h))\frac{1}{\exp(M_h)} \int_{\sX^{m-1}\times A} \prod_{j=0}^{m-1} K(x_j,dx_{j+1})\exp(h(x_m))\\ 
        &\geq\exp(-\lceil (m-1)/2 \rceil \osc(h))\frac{\varepsilon}{\exp(M_h)} \int_{A} \exp(h(x_m))\nu(dx_m)\\ 
        &= \exp(-\lceil (m-1)/2 \rceil \osc(h))\frac{\varepsilon C_h}{\exp(M_h)}\nu_h(A),
    \end{align*}
    where $C_h$ and $\nu_h$ are defined as in~\Cref{thm:inherit_onestep_minor}. This completes the proof.
\end{proof}

{
\begin{remark}
    The bounded oscillation assumption in~\Cref{thm:inherit_multistep_minor} parallels that in~\Cref{thm:HS}. The extension of~\Cref{thm:HS} to~\Cref{thm:HS_ubdd} motivates investigating whether~\Cref{thm:inherit_multistep_minor} can also be extended to the unbounded oscillation assumption under suitable additional conditions. We leave this for future work.
\end{remark}
}

\section{The preconditioned Crank--Nicolson (pCN) algorithm}\label{sec:pcn}

In this section, we apply our theory to the pCN algorithm. Let $\cH$ be a separable Hilbert space equipped with its Borel $\sigma$-algebra $\cB(\cH)$, inner-product $\langle \cdot,\cdot \rangle$, and norm $\|\cdot\|$. Consider a centered Gaussian measure $q = \cN(0,C)$ on $(\cH,\cB(\cH))$, where $C : \cH \rightarrow \cH$ denotes a covariance operator on $\cH$, which is positive, self-adjoint, and
trace-class. Define a probability measure $\pi$ on $(\cH,\cB(\cH))$ as
\begin{equation}
    \pi(du) = \frac{1}{Z}\exp(-\Phi(u))q(du), \label{def:post_law}
\end{equation}
where the potential $\Phi : \cH\rightarrow \bR_+$ is measurable, and $Z = q(\exp(-\Phi))$ is the normalizing constant. We represent $q$ through its Karhunen--Lo\'eve expansion
\begin{equation*}
    q = \cL\left(\sum_{j=1}^{\infty} \lambda_j\xi_j\phi_j\right),\qquad \xi_j \stackrel{\text{i.i.d.}}{\sim} N(0,1),
\end{equation*}
where $\{\phi_j\}_{j\in \bN}$ is an orthonormal basis of eigenvectors of $C$ with eigenvalues $\{\lambda^2_j\}_{j\in\bN}$, and $\cL(\cdot)$ denotes the law of a random variable. We denote by $\Pi_N:\cH \to \cH$ the orthogonal projection onto
$\cH_N := \operatorname{span}\{\phi_1,\ldots,\phi_N\}$ and define $\Phi_N := \Phi|_{\cH_N}$ and $C_N := \Pi_NC\Pi_N|_{\cH_N}$, where $f|_A$ denotes the restriction of $f$ to $A$. The corresponding finite-dimensional approximations of $q$ and $\pi$ are defined by
\begin{align*}
    q_N(du) &= \cL\left(\sum_{j=1}^{N}\lambda_j\xi_j\phi_j\right)(du),\\
    \pi_N(du) &= \frac{1}{Z_N}\exp(-\Phi_N(u))q_N(du), \quad Z_N = q_N(\exp(-\Phi_N)).
\end{align*}
The pCN algorithm is a Metropolis algorithm that targets $\pi$ using a proposal that preserves the Gaussian reference measure; see also~\citet{cotter13mcmc}. The proposal can be viewed as a discretization of an Ornstein--Uhlenbeck (OU) process with invariant measure $q$. Specializing the notation introduced in~\Cref{sec:inheritance} to the pCN setting, let $K$ be the $q$-reversible OU proposal kernel given by
\begin{equation*}
    K(u,\cdot) = \cN(\rho u, (1-\rho^2)C),
    \qquad u \in \cH,
\end{equation*}
for a fixed $\rho \in (0,1)$. The pCN kernel
$P_K$ is defined as in~\eqref{def:mhkernel} with acceptance probability
\begin{equation*}
    \alpha(u,v) = 1 \wedge \exp(\Phi(u)-\Phi(v)),
    \qquad u,v \in \cH.
\end{equation*}
For each $N\in\bN$, we denote by $K_N$ the OU proposal kernel on $\cH_N$ obtained by replacing $C$ with $C_N$, and by $P_{K_N}$ the corresponding pCN kernel with acceptance probability
\begin{equation*}
    \alpha_N(u,v) = 1 \wedge \exp(\Phi_N(u)-\Phi_N(v)),\qquad u,v\in \cH_N.
\end{equation*}
By construction, $K_N$ is $q_N$-reversible and $P_{K_N}$ has invariant
measure $\pi_N$.

The classical drift and minorization method for establishing geometric ergodicity of Markov chains~\citep{meyn2012markov} are not well suited to function-space MCMC algorithms, since the $\psi$-irreducibility of the chain may fail in infinite dimensions. Recall that a Markov kernel $P$ on a measurable space $(\sX,\cF)$ is $\psi$-irreducible if there exists a nonzero measure $\psi$ on $(\sX,\cF)$ such that, for every $x\in \sX$ and $A \in \cF$ with $\psi(A)>0$, we have $\sum_{n=1}^{\infty}P^n(x,A)>0$. In infinite-dimensional settings, this property may fail because transition measures associated with different initial states can be mutually singular. In particular, the pCN kernel $P_K$ is known not to be $\psi$-irreducible under the standard infinite-dimensional setting considered in \citet{hairer14gappcn}. Instead, the spectral-gap methods provide a more appropriate framework for analyzing the geometric convergence of pCN algorithms.

We give certain regularity assumptions on the potential $\Phi$ under which $P_K$ and $P_{K_N}$ have $L^2$-spectral gaps bounded uniformly away from zero. Specifically, we assume that $\Phi$ is locally bounded above and quasiconvex.

\begin{assumption}
    \label{ass:pcn_poincare}
    The potential $\Phi:\cH\to\bR_+$ satisfies the following conditions:
    \begin{enumerate}
        \item[(i)] (bounded above on norm balls) For every $R>0$ there exists a constant $M(R) \geq 0$ such that 
        \begin{equation*}
            \Phi(x) \leq M(R)\qquad \forall x\in B_R(0).
        \end{equation*}
        
        \item[(ii)] (quasiconvex) For every $x,y\in \cH$ and $t\in [0,1]$
        \begin{equation*}
            \Phi(tx+(1-t)y)\leq \Phi(x)\vee\Phi(y).
        \end{equation*}
    \end{enumerate}
\end{assumption}

\Cref{ass:pcn_poincare}(i) also appears in \citet[Assumption~1(a)]{hosseini23gapinfmh}, where the authors note that this condition is often readily verified in Bayesian inverse problems. \Cref{ass:pcn_poincare}(ii) is weaker than convexity of $\Phi$ and is equivalent to convexity of its sublevel sets. Its motivation stems from the observation that a pCN proposal can be represented as a convex combination of the current state of the chain and a Gaussian innovation. In~\Cref{lem:ass_uniform}, we show that~\Cref{ass:pcn_poincare}(i) and (ii) imply that, for a suitable sublevel set of $\Phi$, both the probability that an OU proposal exits the set and the probability of rejection when the chain is outside the set are controlled by Gaussian tail probabilities. In~\Cref{sec:pcn-comparison}, we discuss a class of potentials satisfying~\Cref{ass:pcn_poincare} with an example.

We present our main result in~\Cref{thm:dim_ind_gap_pcn}, which establishes a dimension-independent $L^2$-spectral gap for the pCN kernel under the regularity conditions introduced above.

\begin{theorem}[{Uniform $L^2$-spectral gaps for pCN kernels}]
    \label{thm:dim_ind_gap_pcn}
    Under~\Cref{ass:pcn_poincare}, $P_K$ and $P_{K_N}$ have a common nonzero lower bound on their $L^2$-spectral gaps uniformly in $N$, i.e.,
    \begin{equation*}
        \gap(P_K) \wedge
        \inf_{N\in\bN}
        \gap(P_{K_N}) > 0.
    \end{equation*}
\end{theorem}

For notational convenience, we write $\bN_\infty:=\bN\cup\{\infty\}$ and adopt the convention that every object indexed by $N=\infty$ denotes its full-space counterpart, e.g., $\cH_\infty = \cH$. The strategy of the proof is to apply~\Cref{coro:abs_gap_positivity} to each $P_{K_N}$, while making the lower bounds on the right spectral gaps obtained from~\Cref{thm:HS} or~\Cref{thm:HS_ubdd} uniform in $N$. We gather the ingredients needed to apply~\Cref{coro:abs_gap_positivity} to the pCN kernels in the subsequent lemmas.

We present~\Cref{lem:dim_indep_gap_OU,lem:positivity_OU_proposal} to establish a uniform lower bound on the right spectral gaps of the OU proposal kernels and their positivity.

\begin{lemma}[Uniform $L^2$-spectral gaps for OU proposals]
    \label{lem:dim_indep_gap_OU}
    The OU proposal kernels $K_N$ have a common nonzero lower bound on their $L^2$-spectral gaps uniformly in $N$, i.e.,
    \begin{equation*}
        \inf_{N\in\bN_\infty} \gap(K_N) > 0.
    \end{equation*}
\end{lemma}

\begin{proof}
    For every $N\in\bN_\infty$, the proposal kernel $K_N$ can be viewed as a pCN kernel targeting $q_N$. The corresponding potential $\Phi\equiv0$ satisfies the assumptions of~\citet[Theorem~2.15]{hairer14gappcn}, which yields the desired result.
\end{proof}

\begin{remark}
    A direct derivation of $\gap(K_N)$ in the Hilbert space setting would require additional machinery, such as the notion of Ornstein--Uhlenbeck semigroups~\citep{lunardi2020ornstein}. Since a nonzero lower bound on $\gap(K_N)$ that is uniform in $N$ suffices for the proof of \Cref{thm:dim_ind_gap_pcn}, we do not derive the exact value here.
\end{remark}

\begin{lemma}[Positivity of OU proposal]
    \label{lem:positivity_OU_proposal}
    The OU proposal kernel $K_N$ is a positive operator on $L^2(q_N)$.
\end{lemma}

\begin{proof}
    We consider the case $N=\infty$; the same argument applies to each finite $N \in \bN$, with the corresponding covariance operator $C_N$ and reference measure $q_N$. Define an auxiliary OU proposal kernel
    \begin{equation*}
    K^{1/2}(u,dx) = \cN(dx; \rho^{1/2}u,(1-\rho)C)
    \end{equation*}
    Starting from $u \in \cH$, two successive transitions according to $K^{1/2}$ yield
    \begin{align*}
        v^* &= \rho^{1/2}(\rho^{1/2}u + \sqrt{1-\rho}\xi_1) + \sqrt{1-\rho}\xi_2\\
        &= \rho u + \sqrt{1-\rho}(\rho^{1/2}\xi_1 + \xi_2),
    \end{align*}
    where $\xi_1$, $\xi_2$ are independent with law $\cN(0,C)$. Since $\rho^{1/2}\xi_1 + \xi_2 \sim \cN(0,(1+\rho)C)$,
    \begin{equation*}
        (K^{1/2})^2(u,\cdot) = \cN(\rho u,(1-\rho)(1+\rho)C) = \cN(\rho u,(1-\rho^2)C) = K(u,\cdot).
    \end{equation*}
    Since $K^{1/2}$ itself is an $q$-reversible OU proposal kernel, it is self-adjoint on $L^2(q)$. Therefore, for $f \in L^2(q)$,
    \begin{equation*}
        \langle f,Kf \rangle_{L^2(q)} = \langle f,(K^{1/2})^2f \rangle_{L^2(q)} = \Vert K^{1/2}f\Vert^2_{L^2(q)} \geq 0.
    \end{equation*}
    This proves the claim.
\end{proof}

We next verify~\Cref{ass:uosc1,ass:uosc2,ass:uosc3} for the family of pCN kernels $\{P_{K_N}\}_{N\in\bN_\infty}$, uniformly in $N$. Consistent with the notation used in~\Cref{sec:inheritance}, define
\begin{equation*}
    U_{\tau,N} := \{u\in\cH_N:\Phi_N(u)<\tau\},\quad \tau > 0,\ N \in \bN_\infty.
\end{equation*}
Throughout, $U_{\tau,N}^c$ denotes the complement of $U_{\tau,N}$ in $\cH_N$.

\begin{lemma}[{Uniformity of~\Cref{ass:uosc1,ass:uosc2,ass:uosc3}}]
    \label{lem:ass_uniform}    Under~\Cref{ass:pcn_poincare}, the following  hold:
    \begin{enumerate}
        \item[(U1)] There exists a measurable function $V_0 : \cH \rightarrow [1,\infty)$ and a constant $\tau_0 > 0$ such that
        \begin{equation*}
            \sup_{N\in \bN_\infty}\sup_{u \in U_{\tau_0,N}^c} \frac{K_NV_0(u)}{V_0(u)} < 1\quad\text{and}\quad\sup_{N\in \bN_\infty}\sup_{u \in \cH_N} \frac{K_NV_0(u)}{V_0(u)} < \infty,
        \end{equation*}
        where we use the same notation $V_0$ for its restriction to each $\cH_N$.

        \item[(U2)] $\displaystyle \underset{\tau\rightarrow\infty}{\lim}\sup_{N\in\bN_\infty} \sup_{u\in U_{\tau,N}}K_N(u,U_{\tau,N}^c) = 0$.
        \item[(U3)] $\displaystyle \lim_{\tau\to\infty} \sup_{N\in\bN_\infty} \sup_{u\in U_{\tau,N}^c} \int K_N(u,dv)(1-\alpha_N(u,v)) = 0$.
    \end{enumerate}
\end{lemma}

\begin{proof}
    We consider the Lyapunov function $V_{0}(u)=1+\|u\|^2$ on $\cH$. For each $N\in \bN_\infty$,
    \begin{equation*}
        \tr C_N = \sum_{j=1}^N \lambda_j^2 \leq \sum_{j=1}^\infty \lambda_j^2 = \tr C,
    \end{equation*}
    which implies
    \begin{equation*}
        \sup_{N\in\bN_\infty} \tr C_N = \tr C < \infty.
    \end{equation*}
    Using the second moment formula for Gaussian measures on Hilbert spaces~\citep{DaPrato2006}, we have
    \begin{align*}
        K_NV_0(u) &= 1 + \int_{\cH_N}\Vert \rho u + \sqrt{1-\rho^2}\xi \Vert^2q_N(d\xi) = 1+\rho^2\|u\|^2 +(1-\rho^2)\tr C_N,
    \end{align*}
    and hence
    \begin{equation}
    \frac{K_NV_0(u)}{V_0(u)} =\rho^2+\frac{(1-\rho^2)(1+\tr C_N)}{1+\|u\|^2}\nonumber\leq \rho^2+\frac{(1-\rho^2)(1+\tr C)}{1+\|u\|^2}.\label{eq:quad-drift}
    \end{equation}
    We define $B_{R,N}(0) := \cH_N \cap B_R(0)$ for $R > 0$. Since $\Phi_N=\Phi|_{\cH_N}$, \Cref{ass:pcn_poincare}(i) implies that, for every $R>0$,
    \begin{equation}
        \sup_{u\in B_{R,N}(0)}\Phi_N(u) \le \sup_{u\in B_R(0)}\Phi(u) \le M(R)\qquad \forall N \in \bN_\infty. \label{eq:unifbdd_phiN}
    \end{equation}
    Choose $R > \sqrt{\tr C}$ and $\tau_0>M(R)$. It then follows from~\eqref{eq:unifbdd_phiN} that
    \begin{equation}
        \{u\in\cH_N:\Phi_N(u)\ge\tau_0\}\cap B_{R,N}(0)=\varnothing\qquad \forall N \in \bN_\infty. \label{eq:uniform_sublevel_mass}
    \end{equation}
    Therefore, every $u \in U_{\tau_0,N}^c = \{u \in \cH_N : \Phi_N(u) \geq \tau_0\}$ satisfies $\Vert u \Vert > R$, which gives
    \begin{equation*}
        \sup_{N\in \bN_\infty}\sup_{u \in U_{\tau_0,N}^c} \frac{K_NV_0(u)}{V_0(u)} \leq \rho^2+\frac{(1-\rho^2)(1+\tr C)}{1+R^2} < 1,
    \end{equation*}
    where the last inequality follows from $R^2>\tr C$. In addition,
    \begin{equation*}
        \sup_{N\in \bN_\infty}\sup_{u \in \cH_N} \frac{K_NV_0(u)}{V_0(u)} \leq \rho^2+(1-\rho^2)(1+\tr C) < \infty.
    \end{equation*}
    These bounds establish (U1).
    
    We next verify (U2) and (U3). For $N \in \bN_\infty$ and $u \in \cH_N$, write the pCN proposal $v \sim K_N(u,\cdot)$ as 
    \begin{equation*}
        v = \rho u + (1-\rho)\eta\quad\text{where}\quad \eta = \sqrt{\frac{1+\rho}{1-\rho}}\xi,\quad\text{and}\quad \xi \sim q_N.
    \end{equation*}
    For $R>0$, define the event $A_{R,N}=\{\xi \in B_{R,N}(0)\}$. It follows from~\eqref{eq:unifbdd_phiN} that $\Phi_N(\eta) \leq M_R$ on $A_{R,N}$, where $M_R := M\left(R\sqrt{\frac{1+\rho}{1-\rho}}\right)$. Now, \Cref{ass:pcn_poincare}(ii) gives the bound
    \begin{equation}\label{eq:key}
    \Phi_N(v)\leq\Phi_N(u)\vee\Phi_N(\eta) \leq\Phi_N(u)\vee M_R \qquad\text{on }A_{R,N}.
    \end{equation}
    Choose $\tau > M_R$. On the event $A_{R,N}$, if $u\in U_{\tau,N}$, we then have $\Phi_N(v) \leq \Phi_N(u) \vee M_R < \tau$, and thus $v\in U_{\tau,N}$. In words, whenever $\|\xi\|\le R$, the proposal remains in $U_{\tau,N}$ uniformly over $u\in U_{\tau,N}$. Equivalently, by contraposition, a proposal can leave $U_{\tau,N}$ only if $\|\xi\|>R$. Taking probabilities therefore gives
    \begin{equation}\label{eq:escape-bound}
        \sup_{u\in U_{\tau,N}}K_N(u,U_{\tau,N}^c) \le q_N(B_{R,N}(0)^c).
    \end{equation}
    On the other hand, on the event $A_{R,N}$, if $u\in U_{\tau,N}^c$, we have $\Phi_N(v)\le\Phi_N(u)$ by~\eqref{eq:key} and thus $\alpha_N(u,v)=1$. Therefore, given $\|\xi\|\le R$, every proposal generated from a state $u \in U^c_{\tau,N}$ is accepted. Consequently, a rejection can occur only on $A_{R,N}^c$, and hence
    \begin{equation}\label{eq:rejection-bound}
    \sup_{u\in U_{\tau,N}^c} \int K_N(u,dv)(1-\alpha_N(u,v)) \le q_N(B_{R,N}(0)^c).
    \end{equation}
    By \citet[Lemma 3.1]{hairer14gappcn}, 
    \begin{equation*}
        \sup_{N\in \bN_\infty} q_N(B_{R,N}(0)^c) = q(B_R(0)^c).
    \end{equation*}
    Since \eqref{eq:escape-bound}--\eqref{eq:rejection-bound} hold for every $\tau>M_R$, we obtain
    \begin{align*}
    \limsup_{\tau\to\infty} \sup_{N\in\bN_\infty}
    \left[\sup_{u\in U_{\tau,N}}K_N(u,U_{\tau,N}^c)\vee
    \sup_{u\in U_{\tau,N}^c}
    \int K_N(u,dv)(1-\alpha_N(u,v))\right] \le q(B_R(0)^c).
    \end{align*}
    As $R\to\infty$, the right-hand side converges to zero. Thus, (U2) and (U3) hold.
\end{proof}

Finally, we show that, for all sufficiently large $\tau$, the measure of $U_{\tau,N}$ under $\pi_N$ is bounded away from zero uniformly in $N$. This is required for the localized Holley--Stroock argument for the pCN kernels.

\begin{lemma}
\label{lem:uniform_nonmeasurezeroset}
    There exists a $\tau_0 > 0$ such that, for every $\tau \geq \tau_0$, $\inf_{N\in\bN_\infty}\pi_N(U_{\tau,N}) > 0$.
\end{lemma}

\begin{proof}
    Choose $\tau_0$ as in the proof of~\Cref{lem:ass_uniform}. From~\eqref{eq:uniform_sublevel_mass}, for every $N \in \bN_\infty$, $B_{R,N}(0)\cap U^c_{\tau_0,N} = \varnothing$, which gives $B_{R,N}(0) \subset U_{\tau_0,N}$ and $\sup_{u\in B_{R,N}(0)}\Phi_N(u) \leq \tau_0$. It follows that
    \begin{align*}
        \pi_N(U_{\tau_0,N}) &= \frac{1}{Z_N}\int_{U_{\tau_0,N}}\exp(-\Phi_N(u))q_N(du)\\
        &\geq \int_{B_{R,N}(0)}\exp(-\Phi_N(u))q_N(du)\\
        &\geq \exp(-\tau_0)q_N(B_{R,N}(0)),\\
        &\geq \exp(-\tau_0)q(B_R(0)) > 0,
    \end{align*}
    where the first inequality uses $Z_N = q_N(\exp(-\Phi_N)) \leq 1$, and the third inequality uses $q_N(B_{R,N}(0)) \geq q(B_R(0))$
    {\citep[Lemma 3.1]{hairer14gappcn}}.
    Since $U_{\tau_0,N}\subset U_{\tau,N}$ for every $\tau\geq\tau_0$, the same lower bound holds for all such $\tau$. The lower bound is uniform in $N$, which completes the proof.
\end{proof}

With the above lemmas, we can state the proof of~\Cref{thm:dim_ind_gap_pcn}.

\begin{proof}[Proof of~\Cref{thm:dim_ind_gap_pcn}]
    Since each $K_N$ is positive by~\Cref{lem:positivity_OU_proposal}, the argument of~\Cref{coro:abs_gap_positivity} shows that it suffices to establish a common nonzero lower bound on the right spectral gap $\gap_{\rR}(P_{K_N})$. If $\Phi$ is bounded above, then $\Phi\ge0$ implies that the perturbation $h=-\Phi$ has bounded oscillation. Together with~\Cref{lem:dim_indep_gap_OU}, the conclusion follows directly from~\Cref{thm:HS}. It remains to consider the case in which $\Phi$ is not bounded above. 
    
    We show that the uniform verification of~\Cref{ass:uosc1,ass:uosc2,ass:uosc3} established in~\Cref{lem:ass_uniform} yields Poincar\'e inequalities for $P_{K_N}$ with Poincar\'e constants bounded uniformly in $N$, which gives the desired result. To this end, we retrace the proof of~\Cref{thm:HS_ubdd} while keeping track of the dependence of the relevant constants on $N$. Let
    \begin{equation*}
        \gamma_0 := \inf_{N\in\bN_\infty} \gap_{\rR}(K_N) > 0,
    \end{equation*}
    which is obtained from~\Cref{lem:dim_indep_gap_OU}. Define
    \begin{equation*}
        \varepsilon^\dagger_{\tau} := \sup_{N\in\bN_\infty} \sup_{u \in U_{\tau,N}} K_N(u,U^c_{\tau,N}).
    \end{equation*}
    We denote by $K_{N,\tau}$ the restriction of $K_N$ to $U_{\tau,N}$. By condition (U2) of~\Cref{lem:ass_uniform}, we can choose $\tau_1$ sufficiently large so that $\varepsilon^\dagger_\tau < \gamma_0/2$ for every $\tau \geq \tau_1$. For any $N \in \bN_\infty$ and $\tau \geq \tau_1$, we apply~\Cref{lem:resPI} to obtain 
    \begin{align*}
        \gap_{\rR}(K_{N,\tau}) &\geq \gap_{\rR}(K_N) - \sup_{u \in U_{\tau,N}} K_N(u,U^c_{\tau,N})
        \geq \gap_{\rR}(K_N) - \varepsilon^\dagger_{\tau} > \gamma_0/2.
    \end{align*}
    Since the lower bound is uniform in $N$, we have
    \begin{equation}
        \inf_{N\in \bN_\infty} \gap_{\rR}(K_{N,\tau}) > \gamma_0/2,\qquad \tau\geq \tau_1.\label{eq:unif_PI_res_K}
    \end{equation}
    Therefore, for $\tau \geq \tau_1$, the Poincar\'e constant for $K_{N,\tau}$, $C_{K_{N,\tau}}$, is upper bounded by $2/\gamma_0$.
    
    From condition (U1) of~\Cref{lem:ass_uniform}, there exist $\tau_0 > 0$, $\lambda^\dagger_0 \in (0,1]$, and $b^\dagger_0 < \infty$ such that
    \begin{equation*}
        \lambda_0^\dagger := 1 - \sup_{N\in\bN_\infty}\sup_{u\in U^c_{\tau_0,N}} \frac{K_NV_0(u)}{V_0(u)}\quad\text{and}\quad b_0^\dagger := \sup_{N\in\bN_\infty}\sup_{u\in \cH_N} \frac{K_NV_0(u)}{V_0(u)}.
    \end{equation*}
    For $N \in \bN_\infty$ and $\tau > 0$, define
    $r^\dagger_{\tau,N} := \sup_{u\in U_{\tau,N}^c}
    \int K_N(u,dv)(1-\alpha_N(u,v))$, with $r^\dagger_\tau := \sup_{N\in \bN_\infty} r^\dagger_{\tau,N}$.
    By condition (U3) of~\Cref{lem:ass_uniform}, there also exists $\tau_2 \geq \tau_0$ such that $r^\dagger_{\tau_2} < \lambda_0^\dagger$. 
    
    We set $V = V_0$, $\lambda^\dagger = \lambda^\dagger_0 - r^\dagger_{\tau_2} \in (0,1]$ and $b^\dagger = b_0^\dagger + 1 < \infty$. For each $N\in \bN_\infty$,~\eqref{eq:lyapunov_contraction} and~\eqref{eq:lyapunov_bdd} in the proof of~\Cref{lem:driftinherit} give
    \begin{align*}
        \sup_{u\in U^c_{\tau_2,N}} \frac{P_{K_N}V(u)}{V(u)} & \leq \sup_{u\in U^c_{\tau_2,N}} \frac{K_NV(u)}{V(u)} + r^\dagger_{\tau_2,N} \leq 1 - \lambda^\dagger_0 + r^\dagger_{\tau_2} = 1 - \lambda^\dagger, \quad \text{and} \\
        \sup_{u\in \cH_N} \frac{P_{K_N}V(u)}{V(u)} & \leq \sup_{u\in \cH_N} \frac{K_NV(u)}{V(u)} + 1 \leq b^\dagger_0 + 1 = b^\dagger.
    \end{align*}
    Consequently,
    \begin{equation}
        P_{K_N}V/V \leq (1-\lambda^\dagger) + b^\dagger\ind_{U_{\tau,N}},\qquad \tau\geq \tau_2,\ N\in \bN_\infty. \label{eq:unif_drift_P_K}
    \end{equation}
    
    From~\Cref{lem:uniform_nonmeasurezeroset}, there exists a $\tau > 0$ such that $\tau \geq \tau_1 \vee \tau_2$ and $\inf_{N\in\bN_\infty}\pi_N(U_{\tau,N}) > 0$.  
    Since $h_N := -\Phi_N\leq0$, we set $M_{h_N} := \sup_{u\in \cH_N} h_N(u) \leq 0$. For each $N \in \bN_\infty$, applying~\eqref{eq:HS_ubdd_PI_resMH} with $K = K_{N}$, $h = h_N$, and $U_\tau = U_{\tau,N}$, gives a Poincar\'e inequality for $P_{K_N,\tau}$ with Poincar\'e constant $C_{P_{K_N,\tau}} \leq e^{M_{h_N}+\tau}C_{K_{N,\tau}}$. By~\eqref{eq:unif_PI_res_K} and $M_{h_N} \leq 0$, it is bounded uniformly in $N$:
    \begin{equation*}
        C_{P_{K_N,\tau}} \leq e^{M_{h_N}+\tau}C_{K_{N,\tau}} \leq 2e^\tau/\gamma_0\qquad N\in \bN_\infty,
    \end{equation*}
    or equivalently,
    \begin{equation}
        (2e^\tau/\gamma_0)\cE_{P_{K_N,\tau}}(f,f) \geq \var_{\pi_{N,\tau}}(f)\qquad f\in L^2(\pi_{N,\tau}),\ N\in \bN_\infty. \label{eq:unif_PI_res_P_K}
    \end{equation}
    Finally, applying~\Cref{lem:Lyapunov_Poincare} using~\eqref{eq:unif_drift_P_K} and~\eqref{eq:unif_PI_res_P_K} gives
    \begin{equation*}
        \inf_{N\in\bN_\infty}
    \gap_{\rR}(P_{K_N}) \geq \frac{\lambda^\dagger}{1+2b^\dagger e^\tau/\gamma_0}
    >0,
    \end{equation*}
    which completes the proof.
\end{proof}

\subsection{Comparison to~\citet{hairer14gappcn}}\label{sec:pcn-comparison}

In this subsection, we compare~\Cref{ass:pcn_poincare} with the assumptions on $\Phi$ used in~\citet{hairer14gappcn}. We restate their result in our notation below.

\begin{theorem}[{\citet[Theorems~2.14~and~2.15]{hairer14gappcn}}]
    \label{thm:dim_ind_gap_pcn_hairer}
    Assume that the following conditions are satisfied:
    \begin{enumerate}
    \item[(H1)] There exist $\beta > 0$, $\ell \in (1/2,1)$, $R \geq \bigl(2\beta/(1-\rho)\bigr)^{1/(1-\ell)}$ such that, {with $\omega(x) := \beta x^{\ell}$,
    \begin{equation*}
        \inf_{u_1 \in B_R(0)^c} \inf_{u_2 \in B_{\omega(\Vert u_1 \Vert)}(\rho u_1)}\exp(\Phi(u_1)-\Phi(u_2)) > 0.
    \end{equation*}
    }
    
    \item[(H2)] For every $\kappa > 0$, there exists $C_\kappa < \infty$ such that
    \begin{equation*}
        \sup_{u_1\neq u_2 \in B_r(0)} \frac{|\Phi(u_1) - \Phi(u_2)|}{\Vert u_1-u_2 \Vert} \leq C_\kappa e^{\kappa r},\qquad \forall r>0.
    \end{equation*}
\end{enumerate}
Then, the pCN kernel has a dimension-independent lower bound on the $L^2$-spectral gap.
\end{theorem}

Assumption (H1) requires the acceptance probability to be uniformly bounded away from zero for proposals in a ball centered at the proposal mean when the current state is sufficiently far from the origin. Assumption (H2) controls the growth rate of the Lipschitz constant of the potential, allowing at most sub-exponential growth. Similar, though not identical, assumptions are considered in~\citet[Assumption 1(c)~and~(d)]{hosseini23gapinfmh}. The authors note that their local Lipschitz condition, Assumption 1(d), can be verified in many applications, whereas their acceptance probability condition, Assumption 1(c), may be difficult to verify in many standard inverse problems. Their discussion suggests that verifying Assumption (H1) may also be challenging in practice. By contrast, our~\Cref{ass:pcn_poincare} controls the geometry of the sublevel sets (quasiconvexity) and the values of the potential on norm balls (local boundedness).

To illustrate the distinction, we consider a Bayesian inverse problem whose potential has a local Lipschitz modulus that grows too rapidly to satisfy Assumption (H2), while~\Cref{ass:pcn_poincare} remains satisfied.

\begin{example}[Poisson inverse problem]\label{ex:poisson}
    The following may be viewed as a function-space extension of the model studied by~\cite{arridge18vbpoisson}. Let $u\in\cH$ be the unknown signal, and $G : \cH \rightarrow \bR^m$ be a nonzero bounded linear operator. Conditionally on $u$, the observations $Y_1,\ldots,Y_m$ are independent, with
    \begin{equation*}
        Y_j\sim \operatorname{Poisson}\!\left(\exp(G_ju)\right), \qquad j=1,\ldots,m,
    \end{equation*}
    where $G_j : \cH \rightarrow \bR$ denotes the $j$th component of $G$. Writing $y=(y_1,\ldots,y_m)\in\bN_0^m$ for the observed count vector, define
    \begin{equation*}
        \phi(z) := \sum_{j=1}^{m}\phi_j(z_j),\qquad \phi_j(t) := e^t - y_jt,\qquad z = (z_1,\ldots,z_m)\in \bR^m.
    \end{equation*}
    $\phi(Gu)$ is the negative log-likelihood up to an additive constant. We shift this by $\underline{\phi}$ to define 
    \begin{equation}
        \Phi(u) = \phi(Gu) - \underline{\phi},\qquad u \in \cH. \label{eq:poten_pois}
    \end{equation}
    Then $\Phi(u) \ge 0$, with the target measure $\pi(\cdot)$ defined as in~\eqref{def:post_law}. Here, $\underline{\phi}:=\inf_{z\in\bR^m}\phi(z)>-\infty$, where we use the convention $0\log 0 = 0$. The following proposition shows that this example cannot be treated using the conditions of~\citet{hairer14gappcn}.

    \begin{proposition}
        The potential $\Phi(\cdot)$ in~\eqref{eq:poten_pois} does not satisfy (H2).
    \end{proposition}

    \begin{proof}
    Define the local Lipschitz modulus by
    \begin{equation*}
        L_{\Phi}(r) := \sup_{u_1 \neq u_2 \in B_r(0)} \frac{\lvert \Phi(u_1)-\Phi(u_2)\rvert}
     {\lVert u_1-u_2\rVert}, \qquad r>0.
    \end{equation*}
    Since $G\neq 0$, choose $e\in\cH$ such that $\lVert e\rVert=1$ and $Ge\neq 0$. Replacing $e$ by $-e$ if necessary, we may assume that
    \begin{equation*}
        c:=\max_{1\leq j\leq m}G_je>0.
    \end{equation*}
    Set $u_r=(r/2)e$. Since $u_r,0\in B_r(0)$, for all sufficiently large $r$,
    \begin{align*}
        L_{\Phi}(r) &\geq \frac{\lvert \Phi(u_r)-\Phi(0)\rvert}{\lVert u_r\rVert} = \frac{2}{r}\bigg| \sum_{j=1}^{m} \left(e^{(r/2)G_je} - y_j (r/2) G_je\right) - m\bigg|\\
        &\geq \frac{2}{r} \left( e^{cr/2}-m-\frac{r}{2} \sum_{j=1}^{m}y_j G_je \right)
    \end{align*}
    Thus, for every $0 < \eta < c/2$, $e^{-\eta r}L_\Phi(r)$ tends to infinity as $r\rightarrow \infty$, proving the claim.
    \end{proof}

    Although this potential does not satisfy the conditions of~\citet{hairer14gappcn}, it falls within the scope of~\Cref{thm:dim_ind_gap_pcn}.

    \begin{proposition}
        The pCN algorithm targeting the posterior $\pi(\cdot)$ in~\eqref{def:post_law} with potential $\Phi(\cdot)$ in~\eqref{eq:poten_pois} has a dimension-independent $L^2$-spectral gap. \label{prop:dim_indep_gap_pois}
    \end{proposition}

    \begin{proof}
    By~\Cref{thm:dim_ind_gap_pcn}, it suffices to verify that the potential in~\eqref{eq:poten_pois} satisfies~\Cref{ass:pcn_poincare}(i)--(ii). Since $G : \cH \rightarrow \bR^m$ is bounded, the closure of $G(B_R(0))$ is compact in $\bR^m$ for every $R > 0$. Hence, continuity of $\phi$ implies that $\Phi$ is bounded above on every norm ball, verifying~\Cref{ass:pcn_poincare}(i). Each $\phi_j$ is convex, and thus $\phi$ is convex on $\bR^m$. Since $G$ is linear, $\Phi = \phi \circ G - \underline{\phi}$ is convex and hence quasiconvex, which verifies~\Cref{ass:pcn_poincare}(ii). This completes the proof.
    \end{proof}
    
    We emphasize that the argument in~\Cref{prop:dim_indep_gap_pois} is not specific to the Poisson likelihood. Any continuous, bounded-below, and quasiconvex negative log-likelihood $\phi:\bR^m\to\bR$, composed with a bounded linear forward operator $G:\cH\to\bR^m$, gives rise to a potential $\Phi = \phi\circ G - \underline{\phi}$ satisfying~\Cref{ass:pcn_poincare}.
\end{example}

\section{Discussion}\label{sec:discussion}

We established conditions under which a Metropolis kernel $P_K$ inherits a Poincar\'e inequality from its proposal kernel $K$, when the log-density ratio $h$ is bounded above but unbounded below. Our approach combines a localized Holley–Stroock argument with a Lyapunov drift condition, and we apply it to obtain dimension-independent $L^2$-spectral gaps for the pCN algorithm under quasiconvexity and local boundedness of the potential. Our framework encompasses a broad class of Metropolis–Hastings algorithms with reversible proposal kernels, including the independent Metropolis–Hastings algorithm. It does not cover random-walk Metropolis, whose proposal kernel lacks a stationary distribution, but applies in principle to MALA-type proposals or discretizations of reversible diffusions targeting the reference.

Several directions remain open.
First, we can extend our results to subgeometric convergence rates, using the weak Poincar\'e framework of~\citet{andrieu22pi}. The main obstacle is that Assumption 3 -- vanishing rejection outside superlevel sets -- can fail for contractive proposals such as pCN, since points far from the origin may be mapped closer to it and accepted. Quantifying the rate at which the acceptance probability decays in the tails would allow the localization argument to yield weak Poincar\'e inequalities in place of the standard ones.
Second, extending our results from the right spectral gap to the full $L^2$-spectral gap without assuming positivity of $K$ or resorting to a lazy modification would strengthen the convergence guarantees. Finally, it is of interest to establish the necessity of the assumptions that $M_h < \infty$, and bounded oscillations for the inheritance of uniform ergodicity.

\section*{Acknowledgments}
VR acknowledges support through the National Science Foundation grant DMS-2503118.
\bibliographystyle{abbrvnat}
\bibliography{ref}

\end{document}